\documentclass{JHEP3}

  \usepackage{latexsym,bm,amsmath,amssymb,amsfonts}
  \usepackage{epsfig,graphics,graphicx}
  \usepackage{slashed}
  \usepackage{cite}
  \usepackage[latin1]{inputenc}

  \long\def\comment#1{ }

  \newcommand{\beq}{\begin{eqnarray}}
  \newcommand{\eeq}{\end{eqnarray}}
  
 \def\simge{\mathrel{%
   \rlap{\raise 0.511ex \hbox{$>$}}{\lower 0.511ex \hbox{$\sim$}}}}
\def\simle{\mathrel{
   \rlap{\raise 0.511ex \hbox{$<$}}{\lower 0.511ex \hbox{$\sim$}}}}

\vspace{1.5cm}

\title{Quantitative study of the transverse correlation of soft gluons in high energy QCD}

\author{Emil Avsar\\Institut de Physique Th\'eorique de Saclay,
 F-91191 Gif-sur-Yvette, France\\
        E-mail: \email{Emil.Avsar@cea.fr}}

\author{Yoshitaka Hatta\\Graduate School of Pure and Applied Sciences, University
of Tsukuba, Tsukuba, Ibaraki 305-8571, Japan\\
E-mail: \email
{hatta@het.ph.tsukuba.ac.jp
 }}

\abstract{We examine both analytically and numerically the validity of factorization for
the double dipole scattering amplitude $T^{(2)}$  which appears on the
right hand side of the BK--JIMWLK equation. We demonstrate that, if one
uses a dilute object (e.g., a proton in DIS) as the initial condition,
the correlation in the transverse plane induced by the leading order BFKL evolution
is generally strong, resulting in a violation of the mean field approximation
$T^{(2)} \approx TT$ even at zero impact
parameter
 by a factor ranging from 1.5 to $\mathcal{O}(10)$ depending on the relative
size of the scatterers and  rapidity.
This suggests that, within the experimentally accessible energy interval, the
transverse correlation can significantly
affect the nonlinear evolution of the dipole scattering amplitude. It also
suggests that the nonlinear effects may set in earlier, already in the weak scattering regime.
In the case of the simulation with a running
coupling, the violation of factorization is somewhat milder, but  still noticeable. }

\begin{document}

\sloppy

\section{Introduction}
\setcounter{equation}{0} \label{intro}
High energy scattering near the unitarity limit is a delicate problem which deserves intense
theoretical efforts
 in view of its phenomenological importance at hadron colliders. There is a clear goal of
including nonlinear, saturation
  effects due to the high density of gluons
into the energy evolution of scattering amplitudes,
but a precise determination of when and how these effects should be treated is subject to various
uncertainties depending on the process of interest.
   The problem appears to somewhat simplify if one considers scattering of a small object
  (e.g., a photon at high virtuality in DIS) off a very heavy nucleus where saturation is
important already at relatively
   low energy. For such a process the Balitsky--Kovchegov (BK) equation \cite{Balitsky:1995ub,
Kovchegov:1999yj} is the most commonly studied equation which
   provides a concrete scenario for an approach towards unitarity,
   \begin{eqnarray}
     \partial_Y T_Y(x, y) &=& \frac{\bar{\alpha}_s}{2\pi} \int d^2z \, \mathcal{M}(x,y,z)
     \biggl \{ -T_Y(x,y) + T_Y(x,z) + T_Y(z,y) - T_Y(x,z)T_Y(z,y) \biggr \},\nonumber \\
     \mathcal{M}(x,y,z) &\equiv& \frac{(x-y)^2}{(x-z)^2(z-y)^2}\, , \,\,\,\,\,
     \bar{\alpha}_s \equiv \frac{\alpha_s N_c}{\pi}.
   \label{BK}
   \end{eqnarray}
   Here $T_Y(x,y)$ is the forward amplitude of a dipole of size $|x-y|$ at rapidity $Y$.
 The first three terms on the right hand side contain the BFKL physics \cite{Kuraev:1977fs
,Balitsky:1978ic} while the last term $\sim TT$ ensures that the amplitude saturates the
black disc limit $T\to 1$ which is a fixed point of the equation. Being a closed equation,
\eqref{BK} is amenable to both analytical and numerical approaches, and the properties of the
solution as well as their phenomenological consequences have been discussed extensively over
the past several years (see, reviews \cite{Weigert:2005us,JalilianMarian:2005jf}
and references therein).

    However, it is not often emphasized that the BK equation is a mean field approximation
to a more general equation, namely, the B--JIMWLK equation \cite{Balitsky:1995ub,
Jalilian-Marian:1997gr,JalilianMarian:1997dw,Iancu:2000hn,Ferreiro:2001qy}
 \begin{align}
   \partial_Y T_Y(x,y)=\frac{\bar{\alpha}_s}{2\pi}\int d^2z \, \mathcal{M}(x,y,z)
\biggl \{-T_Y(x,y)+ T_Y(x,z)+T_Y(z,y)-\langle T_Y(x,z)T_Y(z,y)\rangle \biggr \}\,,
   \label{JIM}
   \end{align}
nor is the validity of this approximation fully appreciated. Here the brackets
$\langle\dots\rangle$ denote averaging over the target configurations.
The difference between these two equations is usually
considered to be minor: Although the former obviously discards any kind of existing
correlations in the target wavefunction, this would be justified for a large nucleus at
low rapidity (see, however, \cite{Levin:2003nc}). The subsequent quantum evolution
then generates  correlations which vanish in the large $N_c$ limit,
 \begin{align}
 \langle TT\rangle \approx \langle T \rangle \langle T \rangle + \mathcal{O}\left(\frac{1}
{N_c^2}\right)\,. \label{fact}
 \end{align}
 Indeed, the only existing numerical simulation of the B--JIMWLK equation \cite{Rummukainen:2003ns}
starting from uncorrelated initial conditions shows little difference from the corresponding
solution to the BK equation.

 The purpose of this work is to demonstrate that the factorization (\ref{fact}) is
violated when one considers a dilute target consisting of a few partons (e.g., a proton)
instead of a heavy nucleus as the initial condition. Of course, there is \emph{a priori}
no reason to expect that factorization should work in this case, but there has not been any
quantitative study of the degree of its violation either. For a dilute target, a significant
 part of the rapidity evolution in realistic experiments is in the linear BFKL regime where
the amplitude is rapidly growing but still much less than unity, whereas saturation is
considered to be relevant only
in the late stages of the evolution.\footnote{However, we have found some evidence
that nonlinear effects might set in earlier due to the correlation. See the discussion
in section \ref{sec:results}.} The fluctuations and correlations developed in the
linear regime are so strong that the initial condition that should be used for the nonlinear
evolution equations is a highly nontrivial system of gluons for which the difference between
\eqref{BK} and \eqref{JIM} may turn out to  be crucial, especially for phenomenology.
 Specifically, in the framework of the QCD dipole model ref.~\cite{Hatta:2007fg} found a
power--law correlation
  in the double scattering amplitude\footnote{See also \cite{Braun:2000ua}, though there
seem to be disagreements in the results.}
 \begin{align}
 \langle T(x,z)T(w,y)\rangle \propto  \frac{1}{|z-w|^\gamma} \,, \label{vio}
 \end{align}
under the condition that the distance between the two dipoles are much larger
 than their sizes, $ |z-w| \gg |x-z|, \ |w-y|$. ($\gamma$ is a positive, calculable number
related to the anomalous dimension.) In the exemplary cases studied in \cite{Hatta:2007fg},
this power--law always  leads to a parametrically large ratio
  \begin{align}
  R\equiv \frac{\langle T(x,z)T(w,y)\rangle}{ \langle T(x,z)\rangle \langle T(w,y)\rangle}
\gg 1\,. \label{R}
 \end{align}

Due to a technical reason, in \cite{Hatta:2007fg} it was
not possible to take the interesting
limit $w\to z$ to evaluate $R$ for the `BK configuration', although it was
tantalizing to conclude from (\ref{vio}) that that the correlation would become
even larger in this case.   Here we circumvent this difficulty and
present an analytical insight into the behavior of $R$ as a function of the initial dipole
sizes.

However, analytical calculations are often quite difficult, and one can
usually only deal with special configurations which are set by hand. Besides, for our purpose
it is important to know the actual numerical value of $T$ and $\langle T^2 \rangle$
to make sure that one evaluates $R$ in a regime where the
nonlinear corrections just start to be important.
We will therefore also perform a Monte Carlo (MC) simulation
of the QCD dipole model \cite{Mueller:1993rr} which contains the exact leading order BFKL dynamics.
In this framework one generates a cascade of dipoles keeping track of their
 sizes and positions in the transverse plane. Calculations of  $\langle T^{k}\rangle$
for any $k$, hence $R$, are completely straightforward for arbitrary configurations.
We then compare the numerical results with analytic expectations and find that they
agree satisfactorily.
For zero impact parameter we find that $R$ is much larger than 1 when
the ratio of the projectile and target sizes is either small or large. The
minimum value for $R$ is attained when the projectile and target are of similar
size, and in this case the value of $R$ is around 1.5.
This suggests that, in the leading logarithmic approximation on
which both the BK equation and the dipole
model are based, the replacement $\langle TT\rangle \to \langle T\rangle^2$ is
  not valid for a proton target especially for a small dipole projectile (or in the high--$Q^2$
region of DIS), although it might be safe to do so for a nucleus target.
In
the former case one should rather use the B--JIMWLK equation with a strongly correlated
initial condition, whose asymptotic solution can be different from that
of the BK equation.


The fact that one finds large correlations in the leading order
evolution for a dilute system is consistent with the early studies on
fluctuations in \cite{Salam:1995uy, Mueller:1996te}. In \cite{Salam:1995uy}
it was found that
$\langle T^k \rangle \sim (k!)^2$ (or rather $\langle T^k \rangle \sim  k!\cdot(k+3)!$)
at zero impact parameter.
This implies that, for any $m \leq k$,
\begin{eqnarray}
\frac{\langle T^k \rangle}{\langle T^{k-m} \rangle \langle T^m \rangle}
\sim \biggl ( \frac{k!}{(k-m)!\,m!} \biggr )^2 = \binom{k}{m}^2 \gg 1.
\label{salamfluct}
\end{eqnarray}
Note, however, that
the definition of $\langle T^k \rangle$ in (\ref{salamfluct}) is different
from the one considered in this paper, namely, $\langle T^k \rangle$ appearing in the Balitsky hierarchy whose first equation is (\ref{JIM}). In \eqref{salamfluct}, one evolves the target and the projectile up to some energy, and then calculate the
sum of all events in which there are $k$ simultaneous interactions.
In our case we rather fix $k$ given dipoles in the transverse plane, and then consider their
scattering off some target. Only the latter contains information of the correlation resolved in the
transverse plane.

In \cite{Avsar:2005iz, Avsar:2006jy, Avsar:2007xg} the dipole model has been
modified and extended to include various nonleading effects as well as
saturation and confinement effects during the evolution. Generally speaking, these effects tend to
reduce the correlation.
For example, $\langle T^k\rangle$ as defined in \cite{Salam:1995uy} behaves as
(for $k$ between 5 and 9) $\langle T^k \rangle/
\langle T^{k-1} \rangle \approx 1.2\cdot k$ once the nonleading effects
are included \cite{Avsar:2007xg}. This implies
\begin{eqnarray}
\frac{\langle T^k \rangle}{\langle T^{k-m} \rangle \langle T^m \rangle}
\sim \binom{k}{m}\,,
\end{eqnarray}
and thus the correlation is reduced with respect to (\ref{salamfluct}).
It should, however, be said that the fluctuations
are still very important, and they have for example important consequences on the study
of elastic and diffractive scattering in DIS and
$pp$ collisions \cite{Avsar:2007xg}. In this paper we only show some of the preliminary
numerical results with the
running coupling effect to see if there is a similar suppression of the
correlation, while a detailed study of the various additional
effects is postponed to a future publication.

The paper is organized as follows. In the next section we present
analytical calculations of the double dipole scattering amplitude and the
ratio $R$ for the BK configurations mentioned above. In section
\ref{sec:numeric} we outline our numerical approach to the calculation
of the correlation. The results, including the running coupling case, are
then presented in section \ref{sec:results} where we also make comparison with the analytical expectations.
Finally,  in section \ref{sec:conc} we summarize our results and raise some open questions.

\section{Analytical approach}
\setcounter{equation}{0} \label{Gener}
\subsection{The dipole pair density}

In the dipole model \cite{Mueller:1993rr},
the degree of the two--body correlation in impact parameter space is encoded in the dipole
pair density
\cite{Mueller:1994jq, Mueller:1994gb} whose integral representation
reads (keeping only the zero conformal spin sector) \cite{Peschanski:1997yx, Braun:1997nu}
\begin{eqnarray}
n^{(2)}_Y(x_{01},x_{a_0a_1},x_{b_0b_1})&=&\int d\gamma d\gamma_a d\gamma_b
\frac{1}{2x^2_{a_0a_1}x_{b_0b_1}^2}\int_0^{Y} dy\
e^{\chi(\gamma)y+(\chi(\gamma_a)+\chi(\gamma_b))(Y-y)}
 \nonumber \\ &\times& \int d^2x_\alpha d^2x_\beta
d^2x_\gamma
E^{\gamma}(x_{0\gamma},x_{1\gamma})E^{\gamma_a}
 (x_{a_0\alpha},x_{a_1\alpha})
E^{\gamma_b}(x_{b_0\beta},x_{b_1\beta}) \nonumber \\
&\times& \int \frac{d^2x_2 d^2x_3 d^2
x_4}{x_{23}^2x_{34}^2x_{42}^2} E^{1-\gamma}(x_{2 \gamma},x_{3
\gamma})E^{1-\gamma_a} (x_{2\alpha},x_{4\alpha})
E^{1-\gamma_b}(x_{3\beta},x_{4\beta})\nonumber \\ \label{im}
 \end{eqnarray}
 where $x_{01}=x_0-x_1$ denotes
the coordinate of the parent dipole, and $x_{a_0a_1}=x_{a_0}-x_{a_1}$ and
$x_{b_0b_1}=x_{b_0}-x_{b_1}$ are  those of
 the child dipoles (see, Fig.~\ref{n2}).  We shall use the letter
 $x$ for both  two--dimensional real vectors and their magnitude.  $\chi$ is the BFKL
 eigenvalue \begin{align} \chi(\gamma)=\bar{\alpha}_s
 \bigl(2\psi(1)-\psi(\gamma)-\psi(1-\gamma)\bigr)\,, \end{align} with $\gamma$ being the
anomalous dimension, and  $E$ is the eigenfunction of the SL(2,${\mathbb C}$) group  \begin{align}
&E^{\gamma}(x_{0\gamma},x_{1\gamma})=
\left(\frac{x_{01}}{x_{0\gamma}x_{1\gamma}}\right)^{2\gamma}\,.
\end{align}
 The $\gamma$--integrals are along the imaginary axis. With the usual representation
  $\gamma=\frac{1}{2}+i\nu$, it reads \begin{align} \int d\gamma \equiv \int_{-\infty}^{\infty} d\nu
 \frac{2\nu^2}{\pi^4}\,. \end{align}

\FIGURE[t]{
\includegraphics[height=8.cm]{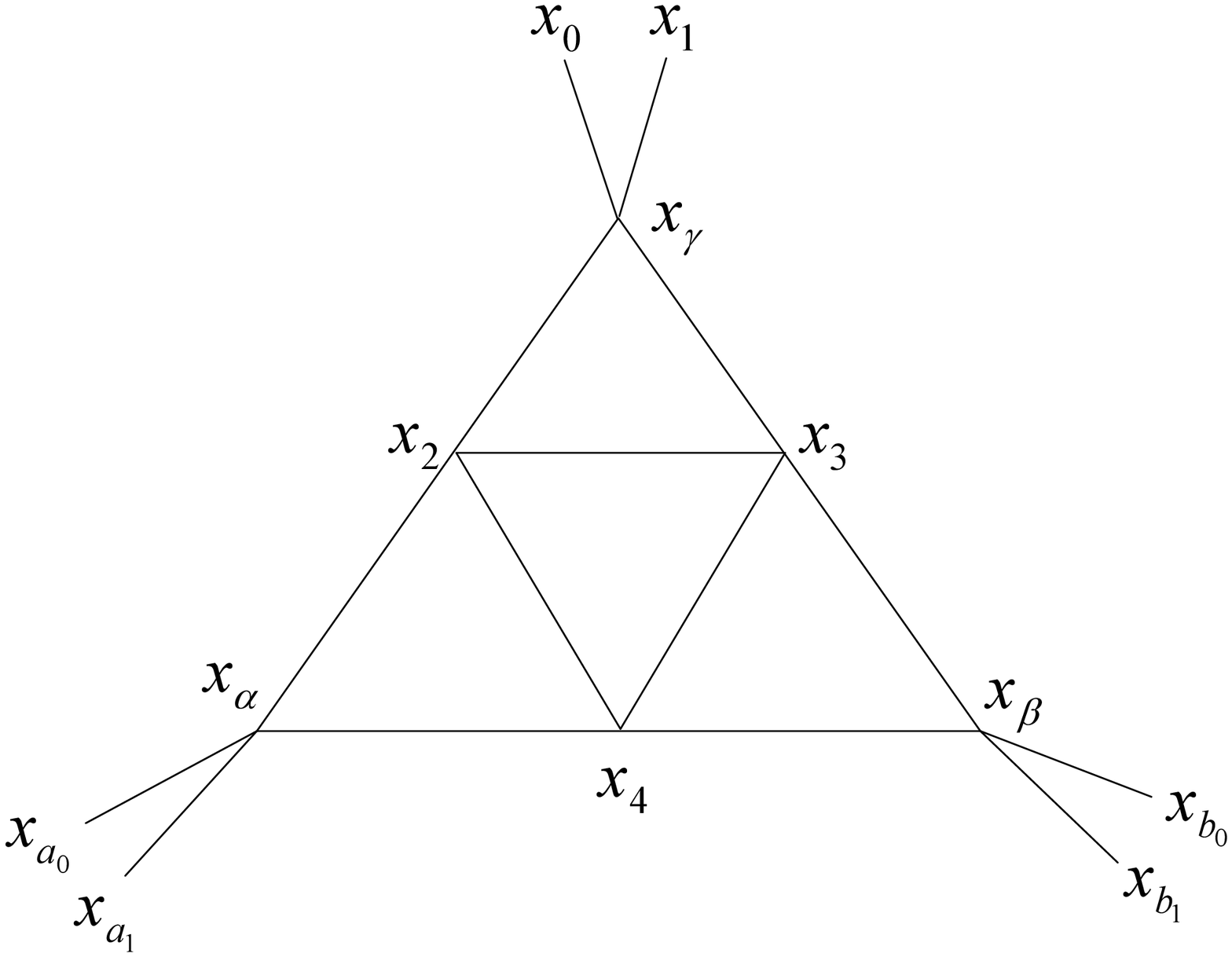}
\caption{ A graphical representation of equation \eqref{im}.
 \label{n2}} }

In ref.~\cite{Hatta:2007fg}, the multi--dimensional integral in (\ref{im}) has been carried
out in the limit
\begin{align} x_{ab}=x_a-x_b\equiv
\frac{x_{a_0}+x_{a_1}}{2}-\frac{x_{b_0}+x_{b_1}}{2} \gg
x_{a_0a_1}, \, x_{b_0b_1}\,. \label{lim} \end{align}  The result shows a power--law
correlation between the two child dipoles. In the case of $x_{01}\gg x_{ab}$, ref.~\cite{Hatta:2007fg}
found
\begin{align} n^{(2)}\sim \left(\frac{x_{01}}{x_{ab}}\right)^{2(2\gamma_a-\gamma)}(n)^2\,,
\label{div} \end{align}
where $n$ is the single dipole density, and $\gamma_a$ and $
\gamma$ are the saddle point values determined from certain conditions. The breakdown of
factorization is carried over to that of the two--dipole scattering amplitude
\begin{align}
T^{(2)}(x_{a_0a_1},x_{b_0b_1})\sim  \left(\frac{x_{01}}{x_{ab}}\right)^{2(2\gamma_a-\gamma)} T(x_{a_0a_1})
T(x_{b_0b_1})\gg T(x_{a_0a_1})T(x_{b_0b_1})\,, \label{nnn}
\end{align}
as already noted in the introduction. (From now on we use the notation
$T^{(2)}$ in place of $\langle T^2 \rangle$.)
On the other hand, the quantity of interest for us is the two dipole scattering amplitude
 for contiguous dipoles,
namely,
\begin{align} x_{a_1}=x_{b_1}\,. \end{align}
 Although it is not legitimate to extrapolate the result (\ref{div})
 to the case $x_{ab}\to 0$, it does suggest that the
correlations would become even larger for such `BK configurations'. (The numerical
evaluation of this case is presented in section \ref{sec:results}.) In this section
we attempt at an analytical evaluation of $n^{(2)}$ for $x_{a_1}=x_{b_1}$ in certain
limits and discuss the behavior of the ratio $R$ defined in (\ref{R}). The result
will be confronted with numerical Monte Carlo simulations in the next section.

\subsection{Calculation of $n^{(2)}$ for contiguous dipoles}

The last line of (\ref{im}) is a known integral whose overall structure is fixed
by conformal symmetry. After performing this integral, the last two lines of (\ref{im}) become
\begin{align} I\equiv f(\gamma,\gamma_a,\gamma_b)\int d^2x_{\alpha} d^2 x_{\beta}
\left(\frac{x_{a_0c}}{x_{a_0\alpha}x_{c\alpha }}\right)^{2\gamma_a} \left(\frac{x_{b_0c}}
{x_{b_0\beta }x_{c\beta}}\right)^{2\gamma_b}
\frac{1}{x_{\alpha\beta}^{2(1+\gamma-\gamma_a-\gamma_b)}} \nonumber \\
 \times \int d^2x_\gamma
\left(\frac{x_{01}}{x_{0\gamma}x_{1\gamma}}\right)^{2\gamma}
\frac{1}{x_{\beta\gamma}^{2(1+\gamma_a-\gamma_b-\gamma)}}
\frac{1}{x_{\gamma\alpha}^{2(1+\gamma_b-\gamma_a-\gamma)}}\,,  \label{1} \end{align}
 where the function $f$--the `triple Pomeron vertex'-- can be found in \cite{Bialas:1997ig,
Korchemsky:1997fy},
and we have already set $x_{a_1}=x_{b_1}\equiv x_c$.

  To make progress we assume that $\gamma_a=\gamma_b$, which is a good approximation
when the configuration of the two child dipoles is more or less
 symmetric. (The saddle points $\gamma_a$ and $\gamma_b$ depend only logarithmically
on dipole sizes.) Then the
 $x_\gamma$ integral can be done \cite{Lipatov:1996ts}
\begin{eqnarray}
\frac{1}{x_{\alpha\beta}^{2(1-\gamma)}} \int d^2x_\gamma
\left(\frac{x_{01}}{x_{0\gamma}x_{1\gamma}}\right)^{2\gamma}
\left( \frac{x_{\alpha\beta}}{x_{\alpha\gamma}x_{\beta\gamma}} \right)^{2(1-\gamma)}
=\frac{1}{x_{\alpha\beta}^{2(1-\gamma)}}\biggl (c_\gamma|\rho|^{2\gamma}|F(\gamma,\gamma,
2\gamma,\rho)|^2 \biggr . \nonumber \\
\biggl . + \,\,(\gamma \leftrightarrow 1-\gamma) \biggr )\,,  \label{two}
\end{eqnarray}
where $F$ is the hypergeometric function, \begin{align}
  c_\gamma=\pi 2^{-4i\nu-1}\frac{\Gamma(\frac{1}{2}+i\nu)\Gamma(-i\nu)}{\Gamma
(\frac{1}{2}-i\nu)\Gamma(1+i\nu)}\,, \label{c} \end{align}  and
 \begin{align} \rho \equiv \frac{z_{01}z_{\alpha\beta}}{z_{0\alpha}z_{1\beta}}\,, \label{rho}
\end{align} is the anharmonic ratio of the four points $(x_0,x_1,x_\alpha,x_\beta)$
($z$ is the complex coordinate
representation of $x$),
see fig.~\ref{after}. The remaining integrals  are difficult to perform in full
generality. As in \cite{Hatta:2007fg}, we shall restrict ourselves to two limiting cases $x_{01}
\to 0$ (small parents) and $x_{01} \to \infty$ (large parents).
 In both limits, $|\rho| \ll 1$, so we may approximate $F(...,\rho)\approx 1$.  The
two terms in
 (\ref{two}) give equal contributions due to the symmetry $\gamma \to 1-\gamma$.
Taking this into account, we can write
 \begin{align}  I= 2c_\gamma f(\gamma,\gamma_a,\gamma_a)  \int \frac{d^2x_{\alpha} d^2
x_{\beta}}{x_{\alpha\beta}^4}  \left(\frac{x_{a_0c}x_{\alpha\beta} }{x_{a_0\alpha }x_{c\beta }}
\right)^{2\gamma_a} \left(\frac{x_{b_0c}x_{\alpha\beta}}{x_{b_0\beta }x_{c\alpha }}\right)^{2\gamma_a}
 \left(\frac{x_{01}x_{\alpha\beta} }{x_{0\alpha }x_{1\beta}}\right)^{2\gamma}\,.  \label{fun}
\end{align}
  The integrand is  a product of anharmonic ratios weighted by the conformally
invariant measure $d^2x_\alpha d^2x_\beta/x_{\alpha\beta}^4$, so it is invariant under
conformal transformations of the external points.
  However, since there are five of them ($x_0$, $x_1$, $x_{a_0}$, $x_{b_0}$ and $x_c$),
conformal symmetry is not strong enough to constrain the solution, and our assumption
$x_{01}\to \infty$ or $x_{01}\to 0$ will be crucial in the following.

\FIGURE[t]{
\includegraphics[height=8cm]{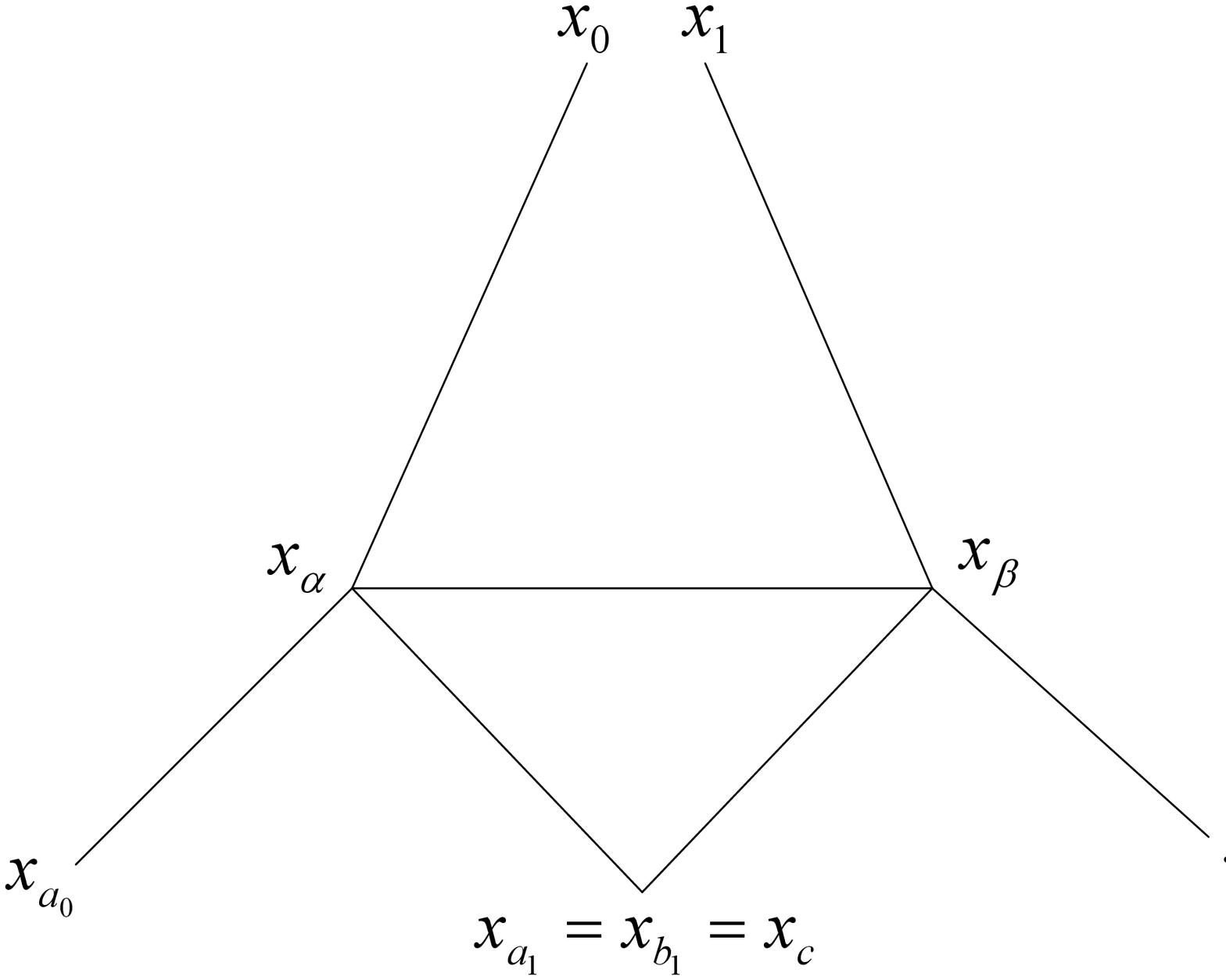}
\caption{ Equation \eqref{im} after integrating over $x_1$, $x_2$, $x_3$ and $x_\gamma$.
 \label{after}} }

    \subsubsection{Large parents}
\label{sec:largep}
Suppose the parent dipole is  large  and the points $x_{a_0,b_0,c}$ are all located near
the center of the parent dipole as illustrated in fig.~\ref{conf}(a). This may be regarded
as a situation relevant to DIS on a hadron at high photon virtuality.  Without loss of
generality, we can set $x_c=0$. The integrand vanishes as $x_{\alpha,\beta}\to \infty$ very
fast, so that a finite region of $x_{\alpha,\beta}$ near the origin is important. Therefore
we may approximate \begin{align} \frac{x_{01}}{x_{0\alpha }x_{1\beta }} \to \frac{4}{x_{01}}\,.
 \end{align} Under this assumption,  (\ref{fun}) takes the form \begin{align} I =2c_\gamma
f(\gamma,\gamma_a,\gamma_a) \left(\frac{4}{x_{01}}\right)^{2\gamma}\int \frac{d^2x_{\alpha} d^2
x_{\beta}}{x_{\alpha\beta}^{4-2\gamma-4\gamma_a}}  \left(\frac{x_{a_0} }{x_{a_0\alpha }x_{\beta }}
\right)^{2\gamma_a} \left(\frac{x_{b_0}}{x_{b_0\beta }x_{\alpha }}\right)^{2\gamma_a}\,. \label{ai}
 \end{align}
\FIGURE[t]{
\includegraphics[height=7.cm]{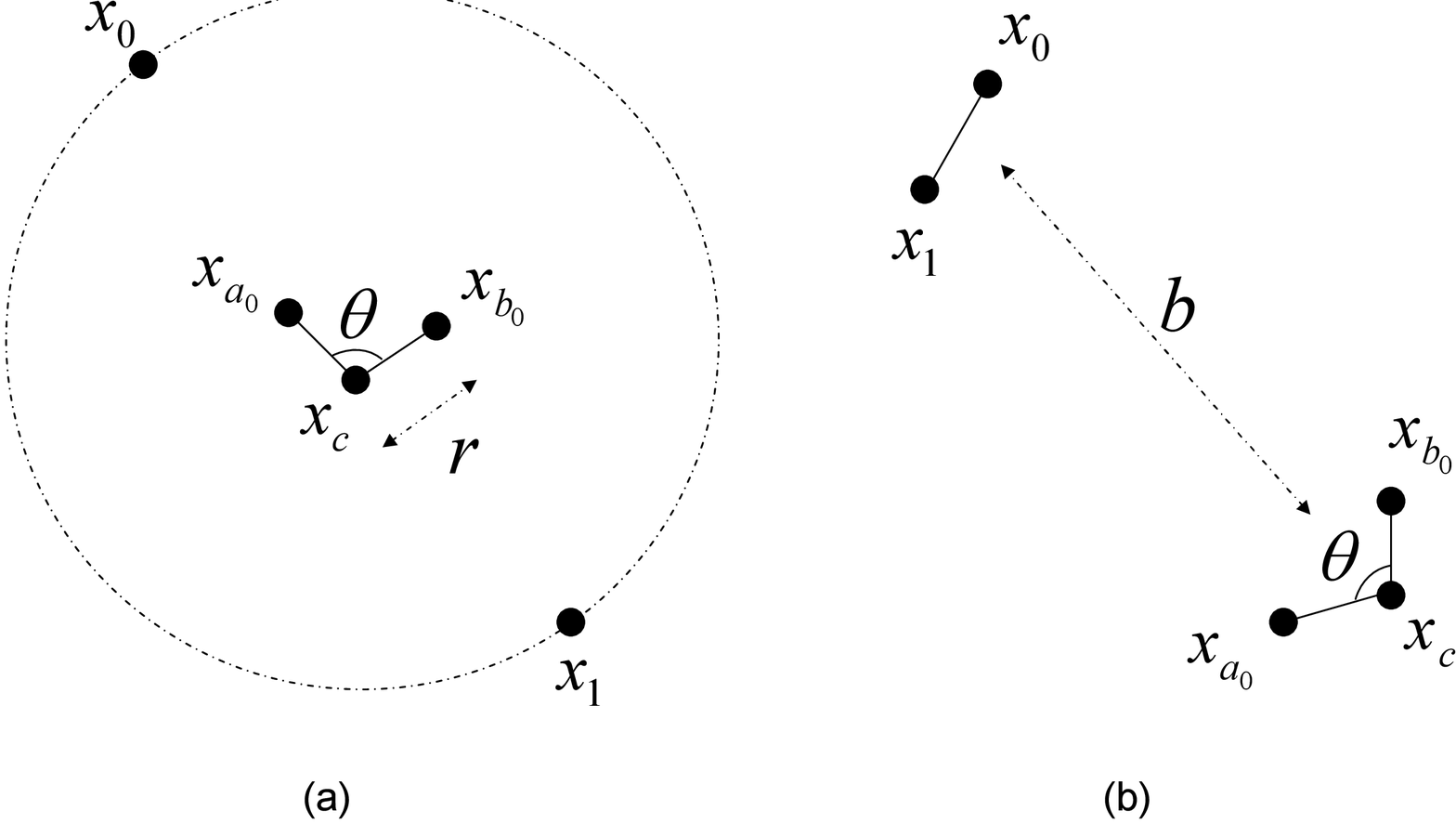}
\caption{ Graphical representation of the 'BK configurations', $(a)$ for
the large parent, small impact parameter case, and $(b)$ for the small parent,
large impact parameter case.
 \label{conf}} }
For simplicity, we assume that the two dipoles have the same size: $x_{a_0c}=x_{b_0c}=r$.
(The region $x_{a_0c}\gg x_{b_0c}$ or $x_{a_0c}\ll x_{b_0c}$ gives a subleading contribution
in the  BFKL or the BK equation, see Section~2.3.)
Writing $z_{a_0}=re^{i\theta_a}$ and $z_{b_0}=re^{i\theta_b}$ and rescaling $x_{\alpha,\beta} \to rx_{\alpha,\beta}$ we get
\begin{eqnarray}
 I&=&2c_\gamma f(\gamma,\gamma_a,\gamma_a)\left(\frac{4r}{x_{01}}\right)^{2\gamma}\int
\frac{d^2x_{\alpha} d^2 x_{\beta}}{x_{\alpha\beta}^{4-2\gamma-4\gamma_a}}  \left(\frac{1 }
{|e^{i\theta_a}-z_\alpha| x_{\beta }}\right)^{2\gamma_a} \left(\frac{1}{|e^{i\theta_b}-z_\beta|x_{\alpha }}
\right)^{2\gamma_a} \label{int}
\nonumber \\
&\equiv& 2c_\gamma f(\gamma,\gamma_a,\gamma_a)\left(\frac{4r}{x_{01}}\right)^{2\gamma}g(\theta)\,,
\end{eqnarray}
where $\theta=\theta_a-\theta_b$ is the relative angle between the two child dipoles.
We have not been able to determine the function $g(\theta)$ for $\theta \neq 0$ in a
closed form ($g(0)$ is a known integral in the conformal field theory literature
\cite{Dotsenko:1984nm, Guida:1996nm,Bondarenko:2007sm}).
But since this function has no singularity and depends
only on the angle, it will  not affect the evaluation of the saddle point below.

Neglecting this angular dependence and other prefactors, we can estimate the two dipole
scattering amplitude as
\begin{align} &T^{(2)}(x_{01},x_{a_0c},x_{b_0c})\sim \alpha_s^2
x_{a_0c}^2x_{b_0c}^2 n^{(2)}(x_{01},x_{a_0c},x_{b_0c})\nonumber \\& \qquad \qquad \qquad \qquad
 \sim \alpha_s^2  \int_0^Y dy \int d\gamma d\gamma_a d\gamma_b \left(\frac{r}{x_{01}}\right)^{2\gamma}
 e^{\chi(\gamma)y+(\chi(\gamma_a)+\chi(\gamma_b))(Y-y)}\,.
\end{align}
After performing the $y$ integral, we get the two contributions
\begin{eqnarray}
\int d\gamma d\gamma_a d\gamma_b\left(\frac{r}{x_{01}}\right)^{2\gamma}
\frac{e^{\chi(\gamma)Y}}{\chi(\gamma)-\chi(\gamma_a)-\chi(\gamma_b)}\,,
\label{contr1}
\end{eqnarray}
and
\begin{eqnarray}
\int d\gamma d\gamma_a d\gamma_b\left(\frac{r}{x_{01}}\right)^{2\gamma}
\frac{e^{(\chi(\gamma_a)+\chi(\gamma_b))Y}}{\chi(\gamma_a)+\chi(\gamma_b)-\chi(\gamma)}\,.
\label{contr2}
\end{eqnarray}
The saddle point for the $\gamma_a$ and $\gamma_b$ integrals in
\eqref{contr2} is simply the BFKL one $\gamma_a=\gamma_b=1/2$, leading to
\begin{eqnarray}
&& \int d\gamma\left(\frac{r}{x_{01}}\right)^{2\gamma}
\frac{e^{2\chi(1/2)Y}}{2\chi(1/2)-\chi(\gamma)}
\sim \left(\frac{r}{x_{01}}\right)^{2\gamma}
e^{2\chi(1/2)Y}\,,
\label{residue}
\end{eqnarray}
where $\gamma$ solves
\begin{align} \chi(\gamma)=2\chi\left(\frac{1}{2}\right)\,, \qquad \gamma\approx 0.82\,
\label{eq:saddle1}.
\end{align}
For the contribution \eqref{contr1} we can use the saddle point for the $\gamma$
integral,
\begin{align}
\chi'(\gamma_s)Y=\ln \frac{x_{01}^2}{r^2}\, ,
\end{align}
and the leading rapidity behavior of this contribution is then given by
\begin{eqnarray}
\left(\frac{r}{x_{01}}\right)^{2\gamma_s}e^{\chi(\gamma_s)Y}.
\end{eqnarray}
As we discuss in section \ref{sec:estimates},  it holds that
$2\chi(1/2) > \chi(\gamma_s)$, \emph{i.e.},
$\gamma_s < 0.82$ for all configurations
we are interested in. (In the limit $Y\to \infty$, $\gamma_s \to 1/2$.) The contribution
which dominates is thus given by \eqref{residue}, and we therefore have
\begin{align} T^{(2)} \sim \alpha_s^2
\left( \frac{r}{x_{01}}\right)^{2\gamma}e^{2\chi(1/2)Y}\,.
\end{align}
On the other hand, the single dipole scattering amplitude is given by
\begin{align}
T(x_{01},r)\sim \alpha_s \left(\frac{r}{x_{01}}\right)^{2\tilde{\gamma}}e^{\chi(\tilde{\gamma})Y}
\,, \label{int22}
\end{align}
where $\tilde{\gamma}$ is the solution to
\begin{align}
\chi'(\tilde{\gamma})Y=\ln \frac{x_{01}^2}{r^2}\,.
\label{cf} \end{align}
Taking the ratio, we arrive at
\begin{align} R\equiv \frac{T^{(2)}}{(T)^2}
\sim \left(\frac{x_{01}}{r}\right)^{2(2\tilde{\gamma}-\gamma)} e^{2(\chi(1/2)-\chi(\tilde{\gamma}))Y}\,.
\label{r1}
\end{align}

Since $2\tilde{\gamma}>1>\gamma$, the first factor is larger than 1 and
predicts that the correlation increases as the asymmetry becomes larger $x_{01}\gg r$.
Since $\chi(\tilde{\gamma})>\chi(1/2)$, the second, exponential factor tends to decrease
the correlation at high values of rapidity.  Comparing this with (\ref{nnn}), we  infer
that $R$ monotonously increases and eventually saturates to the expression (\ref{r1}) as $x_{ab} \to 0$.

\subsubsection{Small parents}
\label{sec:smallp}
Another tractable example is the limit of a small parent dipole $x_{01} \to 0$. In this case
we may approximate $x_{1\beta}\approx x_{0\beta}$, after which the point $x_1$ drops out
from the integral.
  Rewriting
   \begin{align}   I=2c_\gamma f(\gamma,\gamma_a,\gamma_a)\left(\frac{x_{01}x_{a_0b_0} }
{x_{0a_0 }x_{0b_0}}\right)^{2\gamma} \int \frac{d^2x_{\alpha} d^2 x_{\beta}}{x_{\alpha\beta}^4}
 \left(\frac{x_{a_0c}x_{b_0c}x^2_{\alpha\beta} }{x_{a_0\alpha} x_{c\alpha }x_{b_0 \beta}x_{c \beta}}
\right)^{2\gamma_a}
 \left(\frac{x_{\alpha\beta}x_{0a_0}x_{0b_0}}{x_{0\alpha }x_{0\beta}x_{a_0b_0}}\right)^{2\gamma}
 \end{align} we see that, apart from the prefactor, the integrand is conformally invariant,
 so it can be written as  \begin{align} I=2c_\gamma f(\gamma,\gamma_a,\gamma_a)
 \left(\frac{x_{01}x_{a_0b_0} }{x_{0a_0 }x_{0b_0}}\right)^{2\gamma} h\left(\eta, \eta* \right)
\,,  \label{mul} \end{align} where $\eta$ is an anharmonic ratio \begin{align} \eta \equiv
 \frac{z_{a_00}z_{b_0c}}{z_{a_0c}z_{b_00}}\,. \end{align}
  In order to evaluate the function $h$, one can set, using a conformal transformation,
$x_0=\infty$, $x_c=0$, $x_{a_0}=1$
\begin{eqnarray} h(z_{b_0},\bar{z}_{b_0})=\frac{x^{2\gamma_a}_{b_0}}
{(1-x_{b_0})^{2\gamma}}\int d^2x_{\alpha} d^2 x_{\beta} (x_\alpha-x_\beta)^{4\gamma_a+2\gamma-4}
(1-x_\alpha )^{-2\gamma_a}  \nonumber \\
 \times (x_{b_0}-x_{\beta})^{-2\gamma_a} x_{\alpha }^{-2\gamma_a}x_{\beta}^{-2\gamma_a}\,,
\end{eqnarray}
and therefore,
\begin{align} &h(\eta,\bar{\eta})= \frac{|\eta|^{2\gamma_a}}{|1-\eta|^{2\gamma}}
\int d^2x_{\alpha} d^2 x_{\beta} (x_\alpha-x_\beta)^{4\gamma_a+2\gamma-4}(1-x_\alpha )^{-2\gamma_a}
|\eta-z_{\beta}|^{-2\gamma_a}
 x_{\alpha }^{-2\gamma_a}x_{\beta}^{-2\gamma_a}  \nonumber \\
 & \qquad \quad =\left(\frac{x_{01}x_{a_0c} }{x_{0a_0 }x_{0c}}\right)^{2\gamma}
\left(\frac{x_{b_0c}x_{a_00} }{x_{a_0c }x_{0b_0}}\right)^{2\gamma_a} \int d^2x_\alpha d^2x_\beta  \cdots\,.
\end{align}
Remarkably, the same integral as in (\ref{ai}) appears, as a consequence
of the symmetry between the limits $x_{01}\to \infty$ and $x_{01}\to 0$ found
in \cite{Hatta:2007fg}. First consider the case of large impact parameters
$b\equiv |x_{0a_0}|\approx |x_{0b_0}|\approx |x_{0c}| \gg r$ (see, fig.~\ref{conf}(b) and related calculations in \cite{Hatta:2007fg,Bialas:1995yv}). Then  $\eta$
is approximately a phase $\eta\approx e^{i\theta}$ where $\theta$ is the relative angle as before.
We find\footnote{In fact, this result can be reached from (\ref{int}) via a conformal transformation thanks to the conformal invariance of the original integral (\ref{fun}). Consider
a SL(2,${\mathbb C}$) transformation \begin{align}
 z \to z'=\frac{-1}{z-1/b}\,. \end{align}
 Under this, one has $x_{01} \to x'_{01}\approx x_{01}/x_0x_1 \approx 4/x_{01}$, $x_c=0 \to x'_c=b$, $x_{a_0}=r \to x'_{a_0}=b/(1-br)$, and $x_{a_0 c}=r \to r'\approx b^2 r$. Therefore, $4r/x_{01}=x'_{01}r'/b^2$ as expected. Note finally that by definition a conformal transformation does not change the angle $\theta$.
} \begin{align} I \approx 2c_\gamma f(\gamma,\gamma_a,\gamma_a)\left(\frac{x_{01}r}{b^2}\right)^{2\gamma}g(\theta)\,, \end{align}
 and \begin{align}  T^{(2)} \sim \alpha_s^2 \int_0^Y dy \int d\gamma d\gamma_a d\gamma_b
\left(\frac{x_{01}r}{b^2}\right)^{2\gamma} e^{\chi(\gamma)y+(\chi(\gamma_a)+\chi(\gamma_b))(Y-y)}\,.
\end{align}  Again, the saddle points are  given by $\gamma_a=\gamma_b=1/2$,
and we have the pole at $\chi(\gamma)=2\chi(1/2)$.
   On the other hand, the single scattering amplitude at large impact parameter is
\begin{align} T(x_{01},r,b) \sim \alpha_s \left(\frac{x_{01}r}{b^2}\right)^{2\tilde{\gamma}}
e^{\chi(\tilde{\gamma})Y}\,,
\end{align}
where $\tilde{\gamma}$ is the solution to
\begin{align} \chi'(\tilde{\gamma})Y=\ln \frac{b^4}{x_{01}^2r^2}\,.
\label{cf2}
\end{align}
Taking the ratio, we find
\begin{align} R= \frac{T^{(2)}}{(T)^2}\sim
\left(\frac{b^2}{x_{01}r}\right)^{2(2\tilde{\gamma}-\gamma)} e^{2(\chi(1/2)-\chi(\tilde{\gamma}))Y}\,.
\label{r2}
\end{align}
So in this case the correlation $R$ decreases as either $x_{01}$ or $r$
(or both) is increased (keeping $x_{01},r \ll b$).

In order to exhibit a symmetry with respect to the large dipole case, let us look at
the case of small impact parameters, typically, $b\sim r\gg x_{01}$. We
find \begin{align} I \sim \left(\frac{x_{01}}{r}\right)^{2\gamma}\,, \end{align}
while \begin{align} T(x_{01},r,b) \sim \alpha_s\left(\frac{x_{01}}{r}\right)^{2\tilde{\gamma}}
e^{\chi(\tilde{\gamma})Y}\,, \end{align} with $\gamma$ determined from \begin{align}
\chi'(\tilde{\gamma})Y=\ln \frac{r^2}{x_{01}^2}\,,  \label{cf3}
\end{align} so that \begin{align} \label{r3} R\sim \left(\frac{r}{x_{01}}
\right)^{2(2\tilde{\gamma}-\gamma)} e^{2(\chi(1/2)-\chi(\tilde{\gamma}))Y}\,. \end{align}
Compare with (\ref{r1}). As $x_{01}$ increases, while keeping $x_{01}\ll r$, the correlation decreases.
From the limiting behaviors, (\ref{r1}) ($x_{01}\gg r$) and (\ref{r3})
($x_{01}\ll r$), we see that $R$ is enhanced when the asymmetry ($x_{01}$ vs. $r$) is
large, and it presumably takes a minimum value  around $x_{01}\sim r$.

\subsection{Estimates and comments}
\label{sec:estimates}

Regarding the rapidity dependence, we note that $\tilde{\gamma} \to 1/2$
as $Y \to \infty$. Thus for large $Y$, the coefficient multiplying $Y$ in the
exponent in \eqref{r1} and \eqref{r3} tends to zero. For a fixed $Y$, this coefficient
again tends to zero when $x_{01} \to r$, as can be seen from \eqref{cf} and \eqref{cf3}.
Therefore the results \eqref{r1} and \eqref{r3} predict that the correlation $R$ decreases
faster with $Y$ when $x_{01}/r\gg 1$ and $x_{01}/r \ll 1$, while if we
extrapolate our results towards the symmetric
limit $x_{01} \approx r$, we see that $R$ is almost constant in $Y$.

From \eqref{cf} we can guess that $\tilde{\gamma}$ is quite close to $1/2$.
Let us therefore set $\tilde{\gamma} = 1/2 + \epsilon$ and expand
the BFKL eigenfunction to linear order in $\epsilon$. One then finds that
\begin{eqnarray}
\epsilon \approx -\frac{1}{\psi''(1/2)\bar{\alpha}_sY} \mathrm{ln}\frac{x_{01}}{r}
= \frac{1}{14\zeta(3)\bar{\alpha}_sY} \mathrm{ln}\frac{x_{01}}{r}\,,
\label{eq:epsilon}
\end{eqnarray}
where $\zeta(3) \approx 1.2$. If $x_{01}/r=2$ we then find, for $\bar{\alpha}_s=0.2$,
$\epsilon \approx 0.21/Y$,
and thus for $Y=8$ we have $\epsilon \approx 0.03$, while for $Y=12$ we find
$\epsilon \approx 0.02$. For $2(2\tilde{\gamma} - \gamma)$ we then find the
values $0.46$ and $0.43$ for  $Y=8$ and 12 respectively. If instead $x_{01}/r=40$
we find $\epsilon \approx 0.14$, $2(2\tilde{\gamma} - \gamma) \approx 0.91$
and $\epsilon \approx 0.09$, $2(2\tilde{\gamma} - \gamma) \approx 0.73$ for
$Y=8$ and 12 respectively. For this values of $\tilde{\gamma}$ we also note
that the exponent multiplying $Y$ in \eqref{r1} is quite small, for
 $\tilde{\gamma}=0.64$ it is $0.14$ while for $\tilde{\gamma}=0.59$ it is
$0.06$ (all these estimates are valid for $\bar{\alpha}_s=0.2$).

Thus if, for a fixed $Y$, we try to fit $R$ as a function of $x_{01}/r$
using a single effective power, $\omega$, we would expect this fit to give a
too strong increase close to the minimum, $x_{01}/r \sim 1$, whereas it
should give a too slow increase further away from the minimum. As
$2(2\tilde{\gamma} - \gamma)$ varies stronger for smaller $Y$, we would
expect the fit to work better for higher $Y$. We would also expect
$\omega$ to be larger for smaller $Y$.

In the next section we will see that these analytical estimates are
all in quite good agreement with the numerical results. In particular,
the numerical analysis will confirm that the minimum of $R$ (for zero
impact parameter) occurs at $x_{01} \approx r$. Moreover, the estimates for
$\tilde{\gamma}$ given above agree very well with the numerical results,
and also the $Y$ dependence turns out to be correct.


Before moving on to the numerical analysis, we would like to address
one more point. So far we have been able to make analytic estimates
 only for specific configurations. In particular, we assumed that
the dipoles $x_{a_0c}$ and $x_{b_0c}$ are more or less equal in size.
In going from \eqref{JIM} to \eqref{BK}, however, the question
is whether the replacement
\begin{eqnarray}
\int d^2z \, \mathcal{M}(x,y,z) \cdot T_Y^{(2)}(x,z;z,y)
\to \int d^2z \, \mathcal{M}(x,y,z) \cdot  T_Y(x,z)T_Y(z,y)\,,
\label{eq:BKrepl}
\end{eqnarray}
is valid. (We have here returned to the notation used in the introduction
using $x$, $y$ and $z$.)
What we have shown above is that $T^{(2)}(x,z;z,y) \gg T(x,z)T(z,y)$
for some specific regions of $z$, and also for specific relations
between $(x,y)$ and the target, but this is not sufficient to see the integrated
effect of the correlation. Although one can use the MC code
to do the integration over $z$, this can be quite time consuming.
Leaving the numerical integration for future work,
we here crudely identify the configurations which dominate the integral
in \eqref{BK}. Consider the large parent case where
$|x-y| \ll x_{01}$
and assume that $|x-y|$ is smaller than the saturation
length $Q_s^{-1}$. This means that we may set $T(x,y) = (x-y)^2Q_s^2$.
(We could also introduce an anomalous dimension $\gamma \neq 1$ but this
is not essential.) We then divide the integral  into three
regions:
\begin{itemize}\itemsep 0mm
\item Region A:  $|x-z|,\ |y-z| \lesssim |x-y|\,$.
\item Region B:  $|x-y| \lesssim |x-z| \approx |z-y| \lesssim Q_s^{-1}$\,.
\item Region C:  $\, Q_s^{-1} \lesssim |x-z| \approx |z-y|$\,.
\end{itemize}
In region A we have
\begin{eqnarray}
&&\int_A d^2z \frac{(x-y)^2}{(x-z)^2(y-z)^2} \biggl \{ (x-z)^2Q_s^2 + (z-y)^2Q_s^2
-  (x-y)^2Q_s^2 - (x-z)^2Q_s^2\cdot (x-y)^2Q_s^2 \biggr \} \nonumber \\
 && \sim (x-y)^2Q_s^2\,.
\end{eqnarray} (Note that there is no logarithmic singularity at either $z=x$ or $z=y$.)
In region B we instead have
\begin{eqnarray}
&&\int_B d^2z \frac{(x-y)^2}{(x-z)^4}\biggl \{ -  (x-y)^2Q_s^2 + 2(x-z)^2Q_s^2 -
(x-z)^4Q_s^4 \biggr \} \nonumber \\
&&\approx (x-y)^2\int_B d^2z \frac{1}{(x-z)^4}\, 2(x-z)^2Q_s^2 \nonumber \\
&&\sim (x-y)^2\,Q_s^2
\,\mathrm{ln}\frac{1}{(x-y)^2Q_s^2}\,,
\end{eqnarray}
while in region C we have
\begin{eqnarray}
\int_C d^2z \frac{(x-y)^2}{z^4} \biggl \{-  (x-y)^2Q_s^2 + 1 \biggr \}  \sim (x-y)^2\,Q_s^2\,,
\end{eqnarray} where the integral is dominated by the lower limit $|x-z|\sim 1/Q_s$.
Thus for a small projectile which has not yet reached  saturation $|x-y| \ll Q_s^{-1}$,
the dominant contribution comes from region B where we
indeed have $|x-z| \approx |z-y|$. As $|x-y|\to 1/Q_s$, region B shrinks,
and the dominant region is simply  $|x-z|\sim|z-y|\sim |x-y|$.
 Therefore,  we expect that the configurations we are using are relevant,
and the large correlation found there
should survive after integrating over  $z$ in the evolution equation.

\section{Numerical Approach}

\subsection{Outline of the approach}
\label{sec:numeric}

In this section we will perform a numerical analysis to compute the
quantities $T^{(2)}$ and $(T)^2$. This can be done rather easily in a
Monte Carlo implementation of the dipole model, and we will here use the
C++ code developed in \cite{Avsar:2005iz}. The calculation we will perform
is straightforward, no matter which configuration we have.
Recall that the definitions of $T$ and $T^{(2)}$
are
\begin{eqnarray}
T_Y(x,y) &=& \int d^2u\,d^2v\, A_0(x,y|u,v)\, n_Y(u,v)\,,  \\
T_Y^{(2)}(x_1,y_1;x_2,y_2) \!\!&=& \!\! \int d^2u_1\,d^2v_1\,d^2u_2\,d^2v_2 \,
 A_0(x_1,y_1|u_1,v_1)\, A_0(x_2,y_2|u_2,v_2)   n_Y^{(2)}(u_1,v_1;u_2,v_2) \nonumber \\
 && +\int d^2u\,d^2v \,
 A_0(x_1,y_1|u,v)\, A_0(x_2,y_2|u,v)n_Y(u,v)\,, \label{T2def}
\end{eqnarray}
where $A_0$ is the elementary dipole-dipole scattering amplitude. (The second term on
the right hand side of
 \eqref{T2def} represents scattering of two dipoles off the same dipole in the target.)
Starting from any initial dipole distribution, the MC code
evolves the initial state up to a given value of $Y$, after which
one can calculate all possible scatterings between the dipoles.
The Monte Carlo estimate of equation \eqref{T2def} is simply given by
\begin{eqnarray}
T^{(2)}_{MC}(x_1,y_1;x_2,y_2) = \frac{1}{N_{ev}}\sum_{n=1}^{N_{ev}}\,\,
\sum_{i,j \in \Gamma_n} A_0(x_1,y_1|u_i,v_i)\cdot A_0(x_2,y_2|u_j,v_j)\,,
\label{T2MCdef}
\end{eqnarray}
where $\Gamma_n$ is the configuration of the evolved target for the $n$th
event. Writing $\sum_{i,j} = \sum_{i\neq j} + \sum_i$ we see that \eqref{T2MCdef}
contains both contributions in \eqref{T2def}. In writing this formula we only
evolved the target but we can obviously do the computation in any given frame.
Similarly the product $T(x_1,y_1)\cdot T(x_2,y_2)$ is calcuated as
\begin{eqnarray}
T_{MC}(x_1,y_1)\cdot T_{MC}(x_2,y_2) = \frac{1}{N_{ev}}\sum_{n=1}^{N_{ev}}\,\,
\sum_{i \in \Gamma_n} A_0(x_1,y_1|u_i,v_i) \cdot \frac{1}{N_{ev}}\sum_{n=1}^{N_{ev}}\,\,
\sum_{i \in \Gamma_n} A_0(x_2,y_2|u_i,v_i)\,. \nonumber \\
\end{eqnarray}

In the next section we will start by checking the predictions from \cite{Hatta:2007fg}
as stated in equations \eqref{vio} and \eqref{nnn}. As in the analytical approach we 
consider a target which initially consists of a single dipole $(x_0,x_1)$ 
(for the numerical calculation we could start from any configuration if we so wish)
For the configurations in \cite{Hatta:2007fg}, the phenomenologically more relevant
configuration is the one in which the target $x_{01}$ is much larger than
the projectile dipoles. We fix the projectile dipoles to have the same size,
 $r = x_{a_0a_1} = x_{b_0b_1}$ (for the above formulas this means we have 
$x_1 = x_{a_0}, y_1 = x_{a_1}, x_2 = x_{b_0}, y_2 = x_{b_1}$), while the distance
between them, $x_{ab}$, will be varied. 

For the BK configurations, we have $x_2 = y_1 = x_c$, and again we fix the 
two projectile dipoles to have the same size, $r = |x_{a_0}-x_c|=|x_c-x_{b_0}|$. 
The target dipole $(x_0,x_1)$ is placed at
zero impact parameter, as in figure \ref{conf} (a), while its orientation is
chosen randomly for each event. We will always keep $x_{a_0}, x_{b_0}$ and $x_c$ fixed while
we vary $x_{01}$ and the impact parameter.

One technical point is that one has to introduce a cutoff, $\rho$, for the minimal size of
dipoles generated during the evolution since the dipole kernel $\mathcal{M}(x,y,z)$
diverges at $z=x$ and $z=y$. Such a cutoff explicitly breaks conformal symmetry,
and one should therefore ideally choose a cutoff which is much smaller
than the relevant scales (the initial dipole sizes) involved in the process.
On the other hand, simulations with too small values of $\rho$ are very time--consuming.
If one is  studying symmetric collisions
$r \sim x_{01}$, then the choice $\rho=0.01r=0.01x_{01}$ is good enough.
Choosing an even smaller $\rho$  in this case is not useful since
one is then wasting a lot of time to generate many very small dipoles which
do not interact and do not contribute much to the scattering amplitude.
However, here we wish to study the correlation as we vary $x_{01}$, and then the choice of
 $\rho$ is more subtle. For example, for a very asymmetric collision,
say $x_{01} \sim 100 r$,
 $\rho$ has to be much smaller than $0.01x_{01}$ so that we do not
suppress important dipoles with size of order $r$.
 Besides, in the absence of saturation effects, smallness of $\rho$
is also required for the frame--independence of $T^{(2)}$, hence that of $R$.
As a compromise between these requirements (reducing simulation time and
ensuring frame--independence) we shall choose
$\rho(x_{01})=0.05\,r$ throughout. With this choice we confirmed that the
results presented in what follows are reasonably
frame--independent even up to the center--of--mass frame.

\subsection{Results}
\label{sec:results}

As mentioned above we start by checking the results from \cite{Hatta:2007fg}.
 The target
will be fixed at the origin, with random orientation, and the projectile
dipoles are placed symmetrically along the horizontal axis, one on the positive axis
and the other on the negative axis, with random orientations. We choose
$\bar{\alpha}_s = 0.2$ throughout, except in the running coupling case to be presented later.

\FIGURE[t]{
\includegraphics[angle=270, scale=0.7]{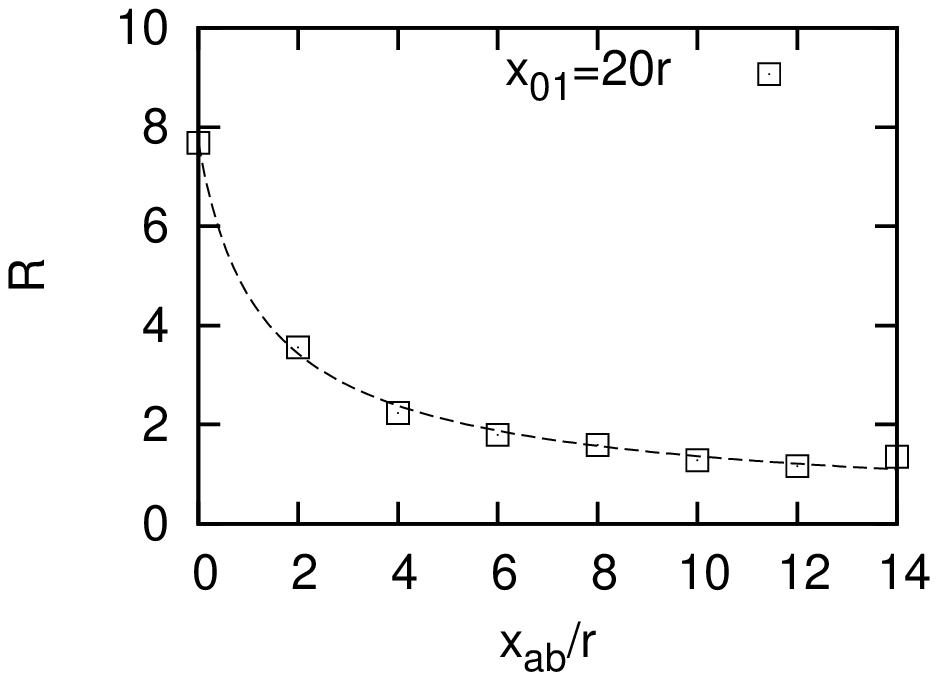}
\includegraphics[angle=270, scale=0.7]{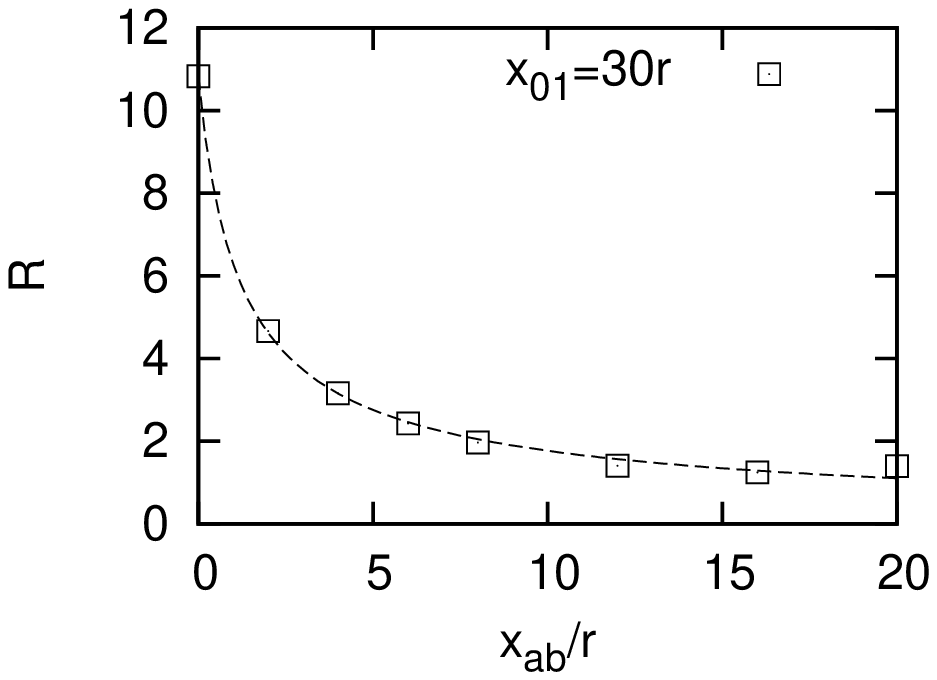}
\caption
{\label{fig:Rresult0} The numerical results for the configurations
described in \eqref{vio} at $Y=10$, and for target of size $x_{01}=20\, r$
(left plot) and $x_{01}=30\, r$ (right plot). The MC results are shown
as squares while the power-like fits to the results are shown as dashed lines.}
}

The results for this configuration are shown in figure \ref{fig:Rresult0}. Here
 we choose $x_{01}=20 \, r$
in the left plot, and $x_{01}=30 \, r$ in the right plot keeping $x_{01}> x_{ab}$.  The former
case would in DIS correspond to a virtuality of $Q^2 \sim 60$ GeV$^2$. 
In both
cases we also show fits of the form $R= \alpha/(x_{ab}+\beta)^\gamma$.
We thus confirm the power--like behavior in \eqref{vio},
and also see that $R$ converges to a finite value as $x_{ab} \to 0$
in agreement with the analytical prediction (\ref{r1}).
For the left plot the fit gives the values $\beta = 0.09$
and $\gamma = 0.70$ while for the right plot we get $\beta = 0.09$
and $\gamma = 0.72$.

\FIGURE[t]{
\includegraphics[angle=270, scale=0.7]{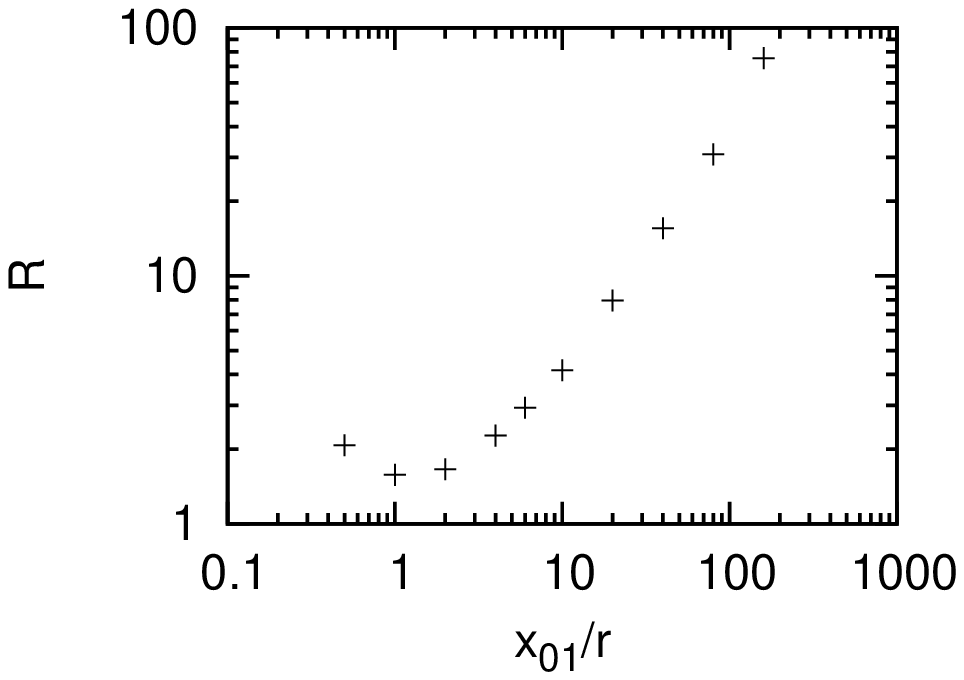}
\includegraphics[angle=270, scale=0.7]{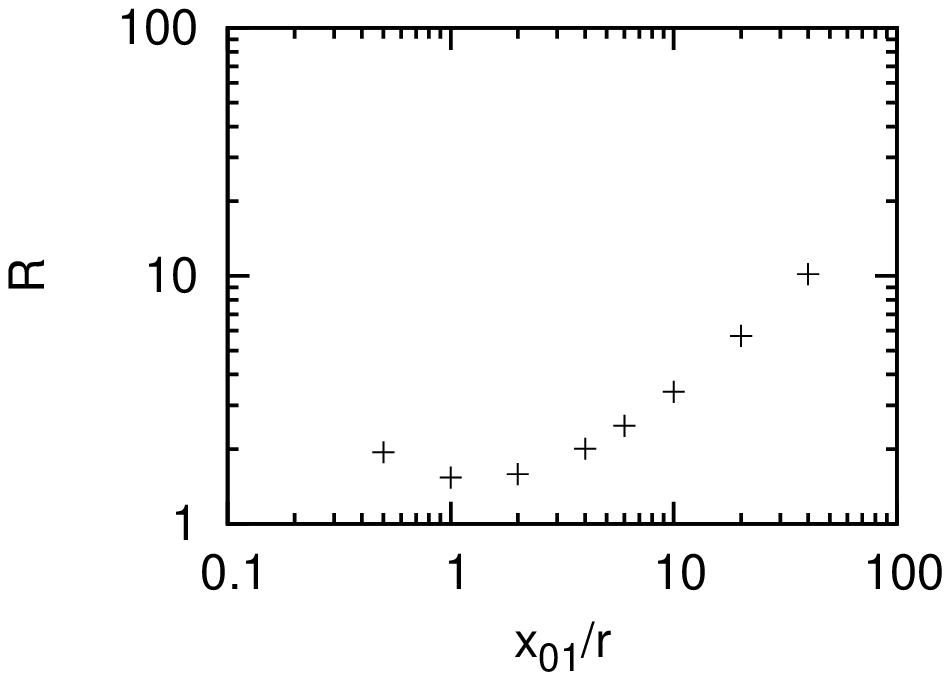}
\includegraphics[angle=270, scale=0.7]{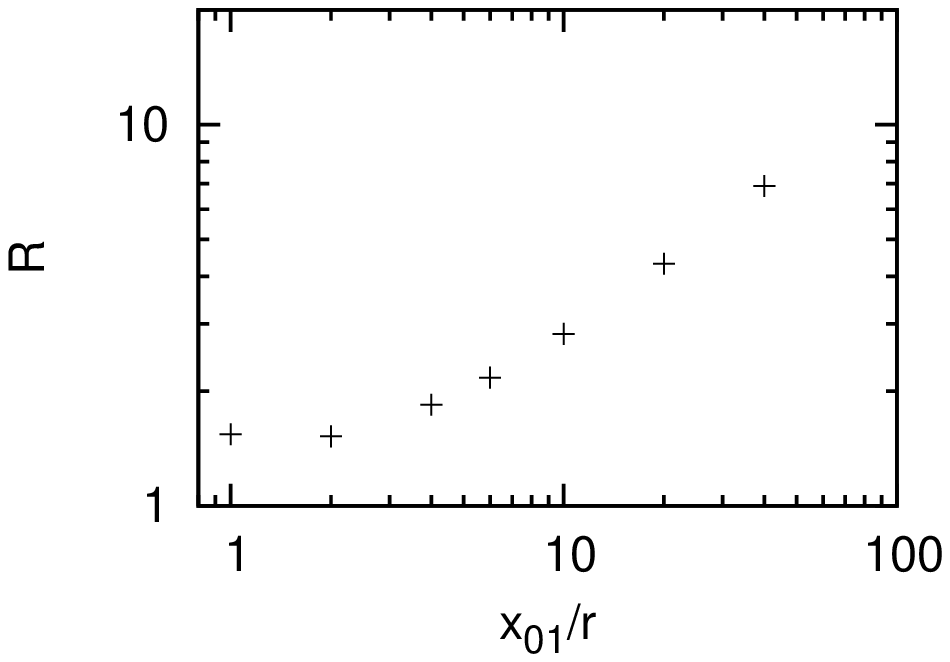}
\caption
{\label{fig:Rresult} The numerical results for $R$ at $Y=6$ (upper left plot), 
  $Y=8$ (upper right plot) and $Y=10$ (bottom plot) at zero impact parameter.
} }

Next we turn to the BK configuration described in the previous section.
In figure \ref{fig:Rresult}, we plot $R$ as a function
of $x_{01}/r$ for $Y=6$, $8$ and $10$, at zero impact parameter. We can  see a
behavior of $R$ consistent with the analytical formulas, equations \eqref{r1} and \eqref{r3}.
The minimum of $R$ indeed occurs at $x_{01} \approx r$ with the minimal value $R\approx 1.5$.
For asymmetric configurations, $R$ can easily reach values of order 10.
 Moreover, the powers extracted from
figure \ref{fig:Rresult}  agree with the expectations
from equations \eqref{eq:saddle1} and \eqref{cf}. For $Y=6$, a fit of the form
$(x_{01}/r)^\omega$ gives the values $\omega=0.55$ in the region $x_{01}/r=1\to 10$,
 $\omega=0.93$ in the region $x_{01}/r=6\to 40$, and $\omega=1.22$ in the region
$x_{01}/r=40\to 200$. These values corresponds to $\tilde{\gamma}=$ 0.55, 0.64
and 0.72 respectively. If we instead calculate the power by calculating 
$\tilde{\gamma}$ using
equation \eqref{eq:epsilon} at the points $x_{01}/r=$ 5, 20 and 120 representing
the three regions above, we find the respective values 0.68, 0.96 and
1.32, in very good agreement with the numerical results.
Similarly, for $Y=8$ we find the values $\omega=$ 0.52 and 0.76
from fits in the first two regions above. This can be compared to the analytical
result which gives $\tilde{\gamma}=$ 0.60 and 0.80.

From our
analysis in the previous section we know that $R$ decreases as $Y$ increases,
and the rate of decrease is larger for asymmetric scattering. This tendency can
be clearly observed, though the ratio $R$ doggedly stays $\gtrsim 1.5$. In the
current simulation we cannot go to larger values of $Y$ because the single
dipole amplitude $T$ for $x_{01}\sim r$ reaches order unity around $Y=10$.
Therefore, in the entire domain of $Y$ values where our approach makes sense,
the mean field approximation $R= 1$ is nowhere valid even in central collisions.
Since this persists up to the onset of the strong scattering regime $T\sim
\mathcal{O}(1)$, it is unlikely that saturation effects immediately wash out
the correlation. Rather, one has to carefully study the effect of correlations
when solving nonlinear equations.

Another, perhaps more striking consequence of the correlation emerging from our analysis
 is that it makes the nonlinear term
 $T^{(2)}$  comparable to $T$ even when $T\ll 1$.
For example, we have $T=0.023$ for $x_{01} = 40\, r$ at  $Y=8$, and in this
case we see from figure \ref{fig:Rresult} that $R=10$. This means that
$T^{(2)} = 0.0056$, and thus $T^{(2)}/T = 0.24$,
so $T^{(2)}$ is  not completely negligible as compared with $T$. For the
more symmetric case $x_{01}=6r$ at $Y=10$ we have $T=0.39$ while $T^{(2)}=0.32$
and $R= 2.2$, see again figure \ref{fig:Rresult}.
For $x_{01} = 40\, r$ and $Y=10$ we instead have $T=0.059$, while
$R=6.9$ and therefore $T^{(2)}/T = 0.41$.
Taken at face value, these estimates suggests that one might have to
include the nonlinear effects in the
 evolution already in the dilute regime  where $T\ll 1$. We did not
include such a back--reaction into our linear dipole evolution, and in this regard our
analysis is not complete. This point certainly deserves  further study.

So far we have studied only configurations with zero impact parameter $b=0$.
At finite impact parameter the correlation becomes larger
as suggested by \eqref{r2}. Of course if we think of
$x_{01}$ as representing the proton radius then one should be careful
in interpreting results for
$b \gg x_{01}$ where confinement effects are certainly important. As a check
of the analytical prediction,
and also for the sake of demonstration, we nevertheless present
some results when $b > x_{01}$. Figure \ref{fig:Rresult3}
shows the $b$ dependence of $R$ for $x_{01}/r=10$ and $x_{01}/r=20$.
We see that $R$ is almost constant as long as $b$ is smaller than $x_{01}$
and that it grows rapidly when $b \gtrsim x_{01}$.
\\

\FIGURE[t]{
\includegraphics[angle=270, scale=0.7]{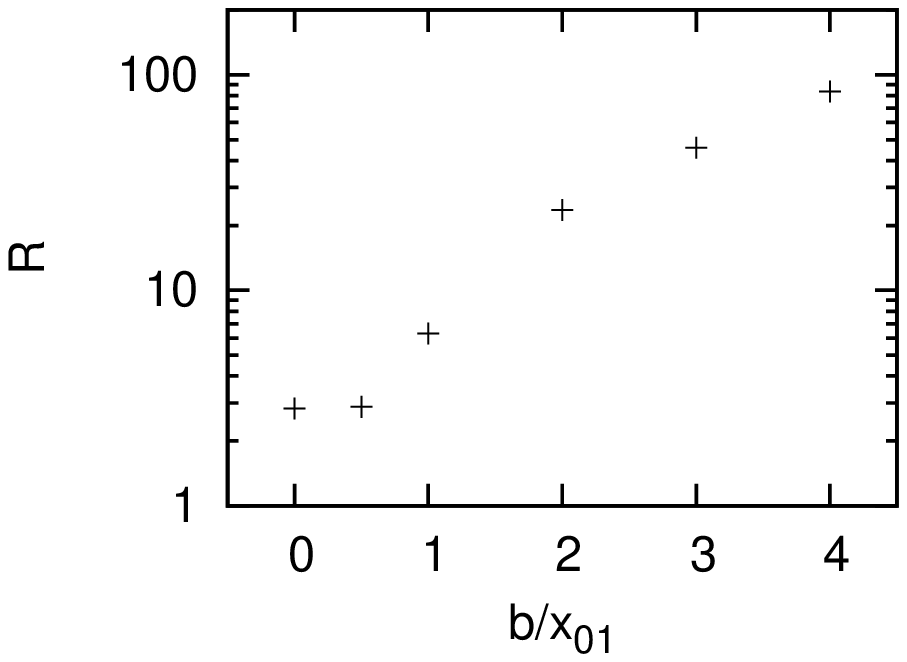}
\includegraphics[angle=270, scale=0.7]{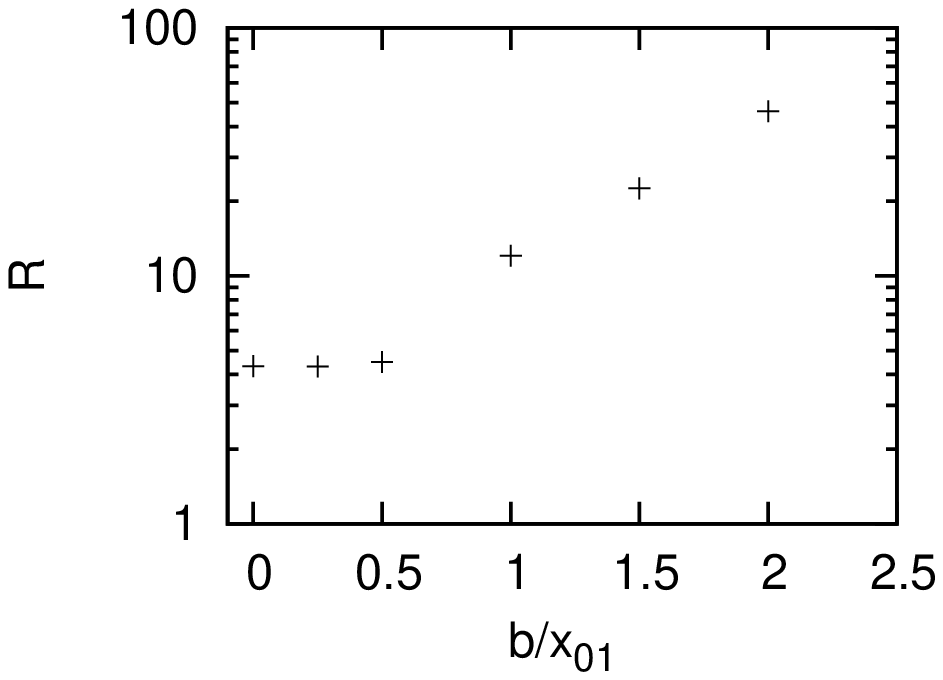}
\caption
{\label{fig:Rresult3} The numerical results for $R$ at nonzero impact parameter
$b\neq 0$, $Y=10$ and $x_{01}=10\, r$
(left plot) and $x_{01}=20\, r$ (right plot).
}}


\emph{Numerical simulation with a running coupling}
\\

One of the non-leading effects which we can easily incorporate into
the numerical simulation is the running coupling as has already been
done in \cite{Avsar:2005iz,Avsar:2006jy, Avsar:2007xg}. Although in
this paper we mainly concentrate ourselves on the fixed coupling case, we would
here like to briefly mention some of the  preliminary results obtained when
the running coupling is used.

Technically, the inclusion of the running coupling is completely straightforward
and we shall use the one-loop expression for $\alpha_s$,
\begin{eqnarray}
\alpha_s(Q^2)=\frac{4\pi}{(\frac{11}{3}N_c-\frac{2}{3}n_f)\,\mathrm{ln}\,(Q^2/\Lambda_{QCD}^2)}
\label{runalpha}
\end{eqnarray}
where we fix $\Lambda_{QCD}=0.22$GeV.
The running coupling enters both in the dipole evolution (as $\bar{\alpha}_s$)
and in the individual dipole-dipole scatterings (as $\alpha_s^2$).
We will  set
 $N_c=3$ and $n_f=3$ as in \cite{Avsar:2006jy, Avsar:2007xg}.

To avoid the IR singularity we shall
freeze the coupling below a minimum scale $Q_{min}$ corresponding to a maximum
dipole size $r_{max}=1/Q_{min}$. As in \cite{Avsar:2007xg}, we choose
$r_{max}=3.5$GeV$^{-1}$. In \cite{Avsar:2007xg}, $\alpha_s$ was evaluated
at the scale $1/Q=\mbox{min}(r,r_1,r_2)$ for the splitting $r\to r_1, r_2$,
and this choice roughly
follows from next--to--leading log (NLL) studies of the dipole evolution
\cite{Balitsky:2006wa,Kovchegov:2006vj}. [See Section~VII of \cite{Balitsky:2008zz}
for a compact discussion.] Thus we continue to use this scale in the evolution of
the dipole cascade. For the dipole--dipole
interaction the correct choice of the scale is more subtle, and we here
use the option described in \cite{Avsar:2007xg}.

\FIGURE[t]{
\includegraphics[angle=270, scale=0.7]{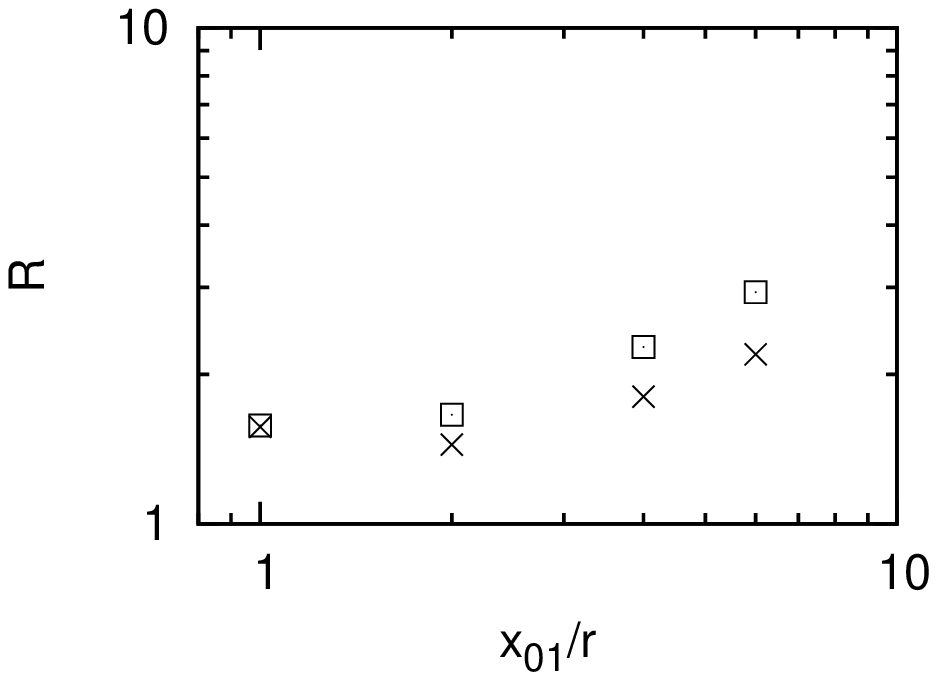}
\includegraphics[angle=270, scale=0.7]{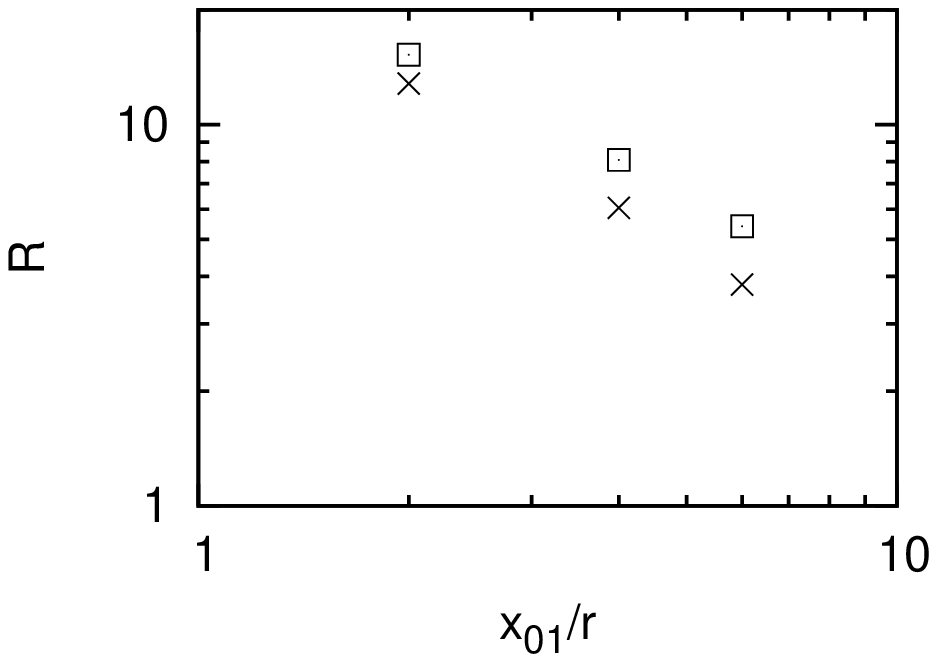}
\caption
{\label{fig:Rrunresult} The crosses are the numerical estimates of $R$
obtained using a running coupling
at $Y=6$, and for $b=0$ (left plot) and $b=5 \, r$ (right plot). The squares
are the corresponding fixed coupling results.}
}

In practice, simulations with the running coupling
are quite time--consuming, and we have therefore not been able to
check as many configurations as in the fixed coupling case. In figure
\ref{fig:Rrunresult} we show the results obtained at $Y=6$ both at zero
(left plot) and nonzero (right plot) impact parameter, together with the
fixed coupling results. We  see that $R$ is somewhat reduced,
but its minimum value is still around 1.5. We also see that the qualitative
behavior of $R$ does not change, the minimum again occurs when
$x_{01} \approx r$ although it is of course difficult to determine the exact
behavior of $R$ since we do not have enough data points. At $Y=8$ for
$b= 0$, we find the value $R=1.5$ at $x_{01}=2\, r$, while in the fixed
coupling case we found $R=1.6$. For $b=5\, r$, $R$ reduces from 11.6
in the fixed coupling case to 9.4 in the running coupling case for
the same configuration.

\section{Conclusions}
\label{sec:conc}

In this paper we have  studied both analytically and numerically the
correlations induced by the leading order BFKL dynamics in the high
energy evolution of a dilute system (such as a proton).
Our main analytical results are given in
equations \eqref{r1}, \eqref{r2} and \eqref{r3}. All these results
indicate that one should expect power--like correlations which lead
to a strong violation of the factorization  $T^ {(2)}
\approx T\cdot T$. The analytical estimates
have been demonstrated to be qualitatively correct by a numerical
analysis with which we have also been able to quantitatively study
the behavior of the ratio $R=T^{(2)}/T^2$. We have found that $R$ is always larger than $\sim 1.5$ and it can easily reach $\sim {\mathcal O}(10)$ when the asymmetry is large.

Physical consequences of the  correlation
remain to be explored. The first and obvious
intuition is that it opens an intriguing possibility of the `grey disc' limit in which
 a scattering amplitude saturates to a value less than 1.\footnote{ Such a possibility
was previously considered in \cite{Levin:2003nc,Janik:2004ts} in the
context of nonlinear equations, although the parameter $R$ in these works was
fixed by some arguments unrelated to the BFKL evolution.}
 \begin{align}
 T \to \frac{1}{R} < 1\,.
 \end{align}
However, since $R$ is not a constant, and the nonlinear equations involve
an integration over the transverse plane
with a nontrivial weight, a more detailed analysis would be required in
order to draw any conclusions.

Another interesting problem is the interplay with the gluon number fluctuation which has
attracted considerable attention lately (see \cite{Dumitru:2007ew} and references therein),
but which has so far mostly been studied in simple toy  models where the transverse dimensions
are suppressed.
Though it typically requires unrealistically large energies  to see the impact of the
gluon number fluctuation on the nonlinear evolution of large nuclei, this is probably
not the case for a dilute target. The BFKL evolution  generates a very strong number
fluctuation as well as the transverse correlation in the dilute regime, and they can
both affect the subsequent nonlinear evolution in significant ways.

There is plenty of room for improvements in the Monte Carlo simulation itself.
In order to make a quantitative prediction for realistic
experiments, one should include various NLL corrections and
saturation effects into the target evolution. They have been incorporated in the dipole
model in \cite{Avsar:2005iz,Avsar:2006jy,Avsar:2007xg}. Among them, we have in this paper
included some results with the running coupling effect.
Since our simulations have been limited in size, it is difficult to determine
the exact behavior of $R$. What we have clearly observed, however, is that $R$
is somewhat reduced from
 the fixed coupling case,
but is still large. This suggests that the large correlation may not be
totally attributed to
conformal symmetry of the leading order BFKL, but rather is a robust feature
of the QCD  evolution in the linear regime.

As mentioned in the introduction we would expect even larger correlations in the
multiple scattering amplitudes $T^{(p)}$  ($p\ge 3$) which enter the Balitsky hierarchy.
In the dipole model, these amplitudes are directly related
to the corresponding multiple dipole distributions $n^{(p)}$ \cite{Peschanski:1997yx,
Braun:1997nu}, but analytical results for them are scarce \cite{Xiao:2007te}. The
numerical evaluation of these amplitudes is straightforward,
although the calculation of $T^{(p)}$ for large $p$ would be  time--consuming
due to the need of good statistics.

\section*{Acknowledgments}

This work was initiated when Y.~H. was a postdoctoral fellow at IPhT, Saclay. He thanks
Riccardo Guida for discussions on complex integrals.

\bibliographystyle{utcaps}
\bibliography{refs}

\end{document}